# Near-perfect fidelity polarization-encoded multilayer optical data storage based on aligned gold nanorods

Linwei Zhu [1†], Yaoyu Cao [2†], Qiuqun Chen [3], Xu Ouyang [2], Yi Xu [2], Zhongliang Hu [3], Jianrong Qiu [3, 4*], and Xiangping Li [2*]

[1] School of Physics and Optoelectronic Engineering, Ludong University, Yantai 264025, China;
[2] Guangdong Provincial Key Laboratory of Optical Fiber Sensing and Communications, Institute of Photonics Technology, Jinan University, Guangzhou,510632, China;
[3] State Key Laboratory of Luminescent Materials and Devices, and Institute of Optical Communication Materials, South China University of Technology, Guangzhou 510640, China;
[4] State Key Laboratory of Modern Optical Instrumentation, College of Optical Science and Engineering, Zhejiang University, Hangzhou 310027, China;
[†] These authors contributed equally to this work.
[*] Correspondence: J R Qiu, E-mail: qjr@scut.edu.cn; X P Li, E-mail: xiangpingli@jnu.edu.cn

**Abstract**: Encoding information in light polarization is of great importance in facilitating optical data storage (ODS) for information security and data storage capacity escalation. However, despite recent advances in nanophotonic techniques vastly enhancing the feasibility of applying polarization channels, the data fidelity in reconstructed bits has been constrained by severe crosstalks occurring between varied polarization angles during data recording and reading process, which gravely hindered the utilization of this technique in practice. In this paper, we demonstrate an ultra-low crosstalk polarization-encoding multilayer optical data storage technique for high-fidelity data recording and retrieving by utilizing a nanofibre-based nanocomposite film involving highly aligned gold nanorods (GNRs). With parallelizing the gold nanorods in the recording medium, the information carrier configuration minimizes miswriting and misreading possibilities for information input and output, respectively, compared with its randomly self-assembled counterparts. The enhanced data accuracy has significantly improved the bit recall fidelity that is quantified by a correlation coefficient higher than 0.99. It is anticipated that the demonstrated technique can facilitate the development of multiplexing ODS for a greener future.

## Introduction

Optical data storage (ODS) with the merits of large capacity, long lifetime, and low energy consumption is considered as one promising way to tackle the challenge of exponentially increasing amount of information [1-4]. This relies heavily on developing effective optical and photonic means to further enlarge the information storage capacity and improve the data qualities, including fidelity, security, and lifetime. On the one hand, applying a variety of new photonic principles, for example by introducing optical super-resolution techniques or deep learning approaches, allows extensive paths to circumvent the light diffraction barrier for increasing information storage density [5-10]. However, such schemes inevitably complicate implementation techniques and thus brings about additional technique limits to the recording media as well as facilities. Meanwhile, multiplexing techniques based on multiple properties of the light field, such as the frequency [11], polarization [12-15], phase angle [16], and optical chirality [17,18] can immensely expand physical dimensions to increase the data storage capacity within the optical diffraction limit space. In addition, this method also enables information encryption for satisfying data safety requirements. As a consequence, information multiplexing in specific ODS media becomes a worthwhile choice in facilitating either big data storage or critical information protection [19-25].

Among various ODS materials that enable multiplexing optical data recording and reading, plasmonic gold nanorods (GNRs) have exhibited exceptional optical and photothermal properties for storing information in multidimensions [19, 26, 27]. The GNRs feature with a longitudinal resonance mode upon the excitation of its surface plasmon oscillation, which enables remarkable linear polarization-based dichroism. This leads to significantly enhanced light absorption for inducing photothermal deformation of GNRs in case that the applied electric field is parallel to longitudinal electron oscillation in GNRs. In this regard, tuning the deformation of GNRs by rotating polarization direction of incident light allows accurate control of two-photon-induced absorption (TPA) and two-photon-induced luminescence (TPL) of GNRs for data encoding. However, these approaches mentioned above in current multiplexing optical data storage (ODS) techniques are mainly based on randomly aligned GNRs. Consequently, in addition to the low utilization efficiency, the strong coupling between randomly orientated GNRs deteriorates the polarization-dependent orthogonality in TPA and TPL [28]. The data fidelity in reconstructed bits has been constrained by severe crosstalk occurring between varied polarization angles during the data recording and reading process, which gravely hindered the utilization of this technique in practice.

Here, we demonstrate ultra-low crosstalk polarization-encoded optical data storage technique for high-fidelity data recording and retrieving by utilizing a nanocomposite film involving aligned gold nanorods (GNRs). The nanocomposite film is composed of nanofibers, which are produced with electrospinning technology [29, 30]. The highly aligned GNRs embedded inside the nanofibers enables polarization-encoded information with high encoding/decoding fidelity and low bit error rate. In order to estimate the fidelity, a correlation coefficient and bit error rate are defined. Using this type of nanocomposite film, we experimentally achieved polarization-encoded multilayer ODS with high fidelity and low crosstalk assembly. Experimental results have evidenced that data recording and retrieving can make a rather high correlation coefficient and extremely low bit error rate by utilizing this type of nanocomposite. The present work highlights a great leap in nanocomposite film material for ODS application and hopefully stimulates the development of new multi-dimensional ODS media.

## Experiment setup and sample preparation

### Experiment setup

The optical setup for realizing the multilayered optical data storage is shown in Fig. 1(a). Both recording and readout were conducted in the same home-built optical setup. For reading, a commercial Ti:sapphire oscillator (Chameleon, Coherent), which emits femtosecond (fs) laser pulses with a duration of 140 fs and a repetition rate of 80 MHz, was used as the laser source. The femtosecond laser was initially expanded and collimated for the following recording and readout processes. The combination of the first half-wave plate (HWP) and the linear polarizer was utilized to adjust the power of the laser. The second HWP placed behind the shutter was used to control the polarization angle relative to the axis of GNRs in the sample. The femtosecond laser was reflected by a dichroic mirror (FF735-Di02-25×36, Semrock) and tightly focused to the sample placed on a three-dimensional stage (P-563.3CD, Physik Instrumente) by an objective (UplanSApo 100×1.40 Oil, Olympus). For readout, the TPL of the sample was excited using low power femtosecond pulses outputted from the same laser. Furthermore, the TPL was collected by the same objective and detected by an avalanche photodiode (APD) (SPCM-AQRH-14-FC, Excelitas Technologies). A short-pass filter was used in front of the APD to filter out the excitation light. To conveniently search the approximate focusing plane of the sample, a LED and a CCD were used, which will be turned off in the recording and readout processes. The home-made software was developed to control the synchronous operation of the shutter, the three-dimensional stage, and the APD acquisition system.

### Preparation of the nanocomposite film containing aligned GNRs

The detail for synthesizing the nanocomposite film with aligned GNRs can be found in Supplementary information (the flow diagram is shown in Figure S1): Firstly, preparing the GNRs solution which can be produced using the same method presented [31]; Secondly, concentrating the prepared GNRs to purify it; and then, mixing the concentrated GNRs into homogeneous precursor solution by adding and fully dispersing the concentrated GNRs to PVA aqueous solution; Thirdly, to produce nanofibers by using the homogeneous precursor solution and an electrospinning facility; Finally, creating the transparent nanocomposite film by utilizing the nanofibers and another material with a refractive index similar to PVA. It should be pointed out that the longitudinal surface plasmon resonance (LSPR) absorption peak of the nanocomposite film experiences a redshift and broadening in comparison to the original GNRs, which can be attributed to the increase of dielectric constant around GNRs in the nanocomposite film [32].

### Recording and readout

Both recording and readout were conducted in the same home-built experimental setup. During recording, the femtosecond laser with the central wavelength corresponding to the LSPR absorption peak of the nanocomposite film is utilized. For sufficiently high laser pulse energy, the GNRs inside of the sample will heat up to above the threshold melting temperature and transform their shape into shorter rods or spherical particles [33]. This leads to a depleted population of GNRs with a certain aspect ratio and orientation, and hence polarization-dependent bleaching occurs in the extinction profile. The threshold of the photothermal melting confines the recording process axially to within the focal volume and provides the ability to record three-dimensionally. The shutter was synchronously controlled by a home-made software with a constant exposure time of 25 ms during the recording process, and it was turned off during the readout. During readout, we used a point scanning method to collect the TPL from the recording region. In the readout processes, the scanning speed is optimized. The moving speed of the three-dimensional position stage was set to 40 μm/s. The shutter was synchronously turned on by the home-made software at the beginning of the scan, and it was turned off when the scan was finished.

## Results and discussion

As an example, a single-layer sample containing aligned GNRs (aspect ratio AR=3.4) was used. The optical microscope image of the sample is shown in Fig. 1(b). From the scanning electron microscope (SEM) image inserted in Fig. 1(b), it is clear to see the collected PVA nanofibers inside of the nanocomposite film exhibits a very high degree of alignment. In addition, the longitudinal axis of GNRs is parallel to the aligned axis of PVA nanofibers, as can be seen from the transmission electron microscope (TEM) image. It should be noted that the consistency or the orientation angle of the nanofibers were influenced remarkably by the concentration of the spinning solution, the rotating speed, and the collecting time. The consistency of the GNRs in the nanofibers can reaches above 85% at the proper concentration of the spinning solution, the rotating speed, and the collecting time [34]. Hence, we can see that a recording material based on aligned GNRs has been fabricated. The LSPR peak of the nanocomposite film sample is located at about 800nm measured by a photometer (see detail in Supplementary information, Figure S2). As shown in Fig. 1(c), we can see that the nanocomposite film has two absorption peaks corresponding to the LSPR and the transversal surface plasmon resonance (TSPR), respectively. When the polarization orientation of the incident light is parallel to the aligned axis of the nanocomposite film, a sharp absorption peak corresponding to the LSPR and a very weak absorption peak corresponding to the TSPR are shown. Conversely, the absorption peak at TSPR becomes obvious, while the absorption peak at LSPR is obviously weakened when the polarization orientation of the incident light is perpendicular to the aligned axis of the nanocomposite film. Ideally, when the polarization orientation of the incident light is perpendicular to the aligned axis of the GNRs, the LSPR peak should disappear completely. The occurrence of the absorption peaks can be attributed to the alignment deviation of a few GNRs. This further proves that the GNRs are arranged along the axes of the nanofibers.

In the experiment, we first verify the alignment angles of gold nanorods obtained from TPL intensity patterns in the scanning microscope image. We retrieve the recordings using LSPR mediated TPL because the nonlinear TPL has a significantly higher angular sensitivity compared with other linear detection methods, such as scattering [35] or extinction [36, 37]. To read the TPL signal, the central wavelength of the femtosecond laser is set at the LSPR absorption peak of the nanocomposite film. When the polarization orientation of the excitation beam rotates around the optical axis, the TPL intensity variation follows a biquadratic cosine function [38], which can be used to determine the alignment angles of the GNRs. Figure 1(d) shows the TPL intensity as a function of the polarization angle. It is shown that TPL intensity reached the maximum when the polarization of the excitation beam was set parallel to the axis of the aligned GNRs, while the minimum is located at the perpendicular direction, which further illuminates that the nanocomposite film possesses a high degree of alignment. The experimental measurement is in good agreement with the theory. It should be pointed out that the TPL intensity signal collected to characterize the polarization dependence in the readout process is not from a single point but an area with a large number of pixels. The detail for detecting the TPL intensity by varying the polarization of the excitation beam can be found in Supplementary information (shown in Figure S3).

Once the alignment angle of GNRs inside of the sample is verified, the polarization orientation of the beam was set parallel to the axis of the aligned GNRs in the recording and readout processes by rotating the second HWP in the experimental setup. The polarization-dependence of the aligned GNRs inside of the sample can effectively reduce crosstalk in both recording and readout processes. In order to confirm the effects of aligned GNRs on the optical data storage, we use the nanocomposite film involving aligned GNRs of AR=3.4 to record and extract a pattern by using the polarization orientation paralleled to the alignment angle of GNRs. Figure 2(a) shows the readout result of the TPL intensity distribution. The recording area is 40 μm × 40 μm in size (50 × 50 pixels). It was recorded and read out with the same polarization orientation (the polarization angle shown in Fig. 1(c)) and the central wavelength $\lambda$ = 800 nm. The recording and readout power are 1000 μW and 50 μW, respectively. Basically, we use the correlation coefficient and bit error rate to characterize the quality of a pattern extracted by detecting the TPL intensities of all the information units. The extracted pattern can be binarized by choosing an appropriate threshold intensity, as shown in Fig. 2(b). The correlation coefficient between the binarized pattern and the original one (shown in Fig. 2(c)) can be defined as follows [27]:

$$c = \frac{\sum_m \sum_n (A_{mn} - \bar{A})(B_{mn} - \bar{B})}{\sqrt{\left[\sum_m \sum_n (A_{mn} - \bar{A})^2\right]\left[\sum_m \sum_n (B_{mn} - \bar{B})^2\right]}} \quad (1)$$

where $\bar{A} = \sum_m \sum_n A_{mn}/(mn)$, $\bar{B} = \sum_m \sum_n B_{mn}/(mn)$. Here, $A_{mn}$ and $B_{mn}$ represent the intensities of individual pixel-units ($m$, $n$) while $\bar{A}$ and $\bar{B}$ denote the averaged intensities of all the pixels in the readout binarized and original patterns, respectively. The bit error rate (BER) can be defined as:

$$e = \frac{\sum_m \sum_n |A_{mn} - B_{mn}|}{N} \quad (2)$$

where $|A_{mn} - B_{mn}|$ denotes the error bit between the original data and the extracted binarization data of all the information units. $N$ is the total number of information units.

For the readout pattern shown in Fig. 2(a), the distribution of the TPL intensities of all the pixels is shown in Fig. 2(d), in which the statistical results of the TPL intensity collected in each pixel shows two Gaussian distributions with different average intensities, corresponding to the information units without and with the irradiation of femtosecond laser pulses, respectively. By selecting suitable thresholds for the normalized TPL intensity between the two Gaussian distributions, the two types of information units can be discriminated and binarized. Figure 2(b) shows the binarized pattern at the threshold intensity $I_{th} = 0.37$. The correlation coefficient can be deduced by Eq. (1) after the binarization of the readout pattern. We can see that near-perfect fidelity (correlation coefficient $c = 0.997$) of information storage has been realized. These results demonstrate that it is feasible to record and read out the data information in the nanocomposite nanofibers film, which can be used for optical data storage with high-quality and low crosstalk (bit error rate $e=0.02\%$, see Supplementary information in Figure S4).

Generally, the quality of a recorded pattern characterized by the correlation coefficient can be improved by increasing the recording energy. However, the crosstalk between different recording channels will become larger as the recording energy increase, which results in the reduction of storage capacity and possibly the damage of storage media [39]. Therefore, how to optimize recording energy becomes an issue in the multiplexed optical storage. In order to determine the optimal recording energy of the femtosecond laser, we study the relationship between the correlation coefficient (combined with bit error rate) and the various recording powers.

Figure 3 shows the correlation coefficient and the bit error rate of different TPL images as a function of recording power. In this example, we use the nanocomposite sample with the LSPR peak at about 800 nm. In both recording and readout, the central wavelength of the femtosecond laser is set as 800 nm. In recording, the laser power changes from 300 μw to 3000 μW. The polarization of the recording laser is parallel to the aligned orientation of the GNRs embedded in the film. In the readout, the power of the laser is 80 μw and remains unchanged in the whole readout process. We choose two polarization orientations to read out the TPL signal, corresponding to parallel and perpendicular to the aligned orientation of the GNRs, respectively. From Fig. 3, we can see that as the recording power increases to 600 μW in the case of the polarization is parallel to the aligned orientation of sample, the correlation coefficient increases sharply and then slowly increase to a certain value when the recording power exceeds 600 μW. This demonstrates the recording power has to reach a threshold to achieve a high correlation coefficient and a low bit error rate. Although the correlation coefficient can be improved by increasing the recording power, the crosstalk will occur, and the correlation coefficient will decrease as the recording power increases.

As shown in Fig. 3, when the polarization of the reading laser is perpendicular to the aligned orientation, the correlation coefficient increases sharply when the recording power is above 1800 μW, meanwhile the bit error rate reduces gradually. That is because the sample is damaged and loses the polarization dependency by using high laser power. When the recording power increases a high value, the melting of these GNRs will lead to crosstalk at other polarization angles and decrease the fidelity of extracted information. Hence, high fidelity of information encoding can be realized by choosing a suitable recording power range from 600 μW to 1800 μW.

Furthermore, we demonstrate the multilayer polarization-encoded ODS in Fig. 4. In the experiment, two layers of nanocomposite films with aligned GNRs (PVA-GNRs-3.4-PVP) were prepared by the thermocompression-immersion-drying method. The two orientations of nanofibers inside of the nanocomposite films can be observed in the micrographic image, as shown in Fig. 4(a). Figure 4(b) shows a sketch of the multilayered sample. In this two-layer sample, the spacing and the orientations of aligned GNRs in the two layers can be determined by reading the TPL at various depths and polarizations. Figure 4(c) shows the TPL intensity signal changed as a function of Z scan depth at two polarization angles. The space between the two layers is about 4 μm. To verify the alignment angles of GNRS obtained from their intensity patterns in the TPL scanning images, we performed the TPL polarization sensitivity experiment at the corresponding layer, as shown in Fig. 4(d). It is shown that TPL intensity reaches a maximum when the polarization of the excitation beam is set to 20° and 60° at the corresponding layers, respectively. Therefore, after we confirm the polarization of the recording light, by using different combinations of multilayers and polarizations (i.e., layer 1 and polarization P1, layer 2 and polarization P2), we can realize the polarization-encoded multilayered ODS in the sample with aligned GNRs, as schematically shown in Fig. 4(b). Using the corresponding polarization at each layer, one can retrieve the encoded information in the same spatial region, as shown in Fig. 4(e). The statistical results of the TPL signals under different readout polarizations are also given in Fig. 4(f). The high-quality data storage with near-perfect fidelity (correlation coefficient c > 0.99) was realized in the multilayers. As can be seen from these results, very small crosstalk among different polarization channels was observed (bit error rate e≈ 0.1%, see Supplementary information in Figure S5), validating the possibility of realizing polarization-encoded multilayers ODS with near-perfect fidelity utilizing the materials of the aligned GNRS. It should be pointed out that there are still some bit errors in the readout process, which can be attributed to the inhomogeneity of the sample. For the inhomogeneity of the sample, it may be due to the low doping concentration of GNRs in the single-layer sample, or the distribution of GNRs is not uniform. Another possible reason may be the GNR orientation in nanofibers is not uniform. Besides, the nonuniformity of the PVP during the filling process can also lead to bit errors in the data readout. Thus, further improvement in the preparation process of the nanocomposite film is the key to realize the aligned GNRs successfully for information storage. By contrast, the fidelity of data encoding in the ODS medium with randomly aligned GNRs is less than that in the material of aligned GRNs, and the bit error rate readout from disorder GNRs is more than ten times higher as that from aligned GNRs (shown in Figure S6 of Supplementary information).

In addition, compared with the randomly aligned GNRs for polarization-encoded multilayered ODS, the prepared nanocomposite film composed of aligned GNRs will be highly beneficial in reducing optical losses and efficiently delivering beam energy down to bottom layers. A significant improvement in the transmitted recording beam intensity by multilayers containing aligned GNRs can be achieved at all layers (shown in Figure S7 of Supplementary information). Therefore, this property of polarization-encoded TPL readout in progressively twisted aligned GNRs arrays through the multilayers can further reduce the loss in multilayer ODS and simultaneously maintain near-perfect fidelity for encoding information, which is previously impossible with randomly aligned GNRs.

## Conclusions

In summary, we have presented an electrospinning technology to prepare ODS recording media composed of a nanocomposite film doped with aligned GNRs. We achieved experimentally multilayers ODS with near-perfect fidelity and low bit error rate by utilizing the nanocomposite film supporting both polarization and multi-dimension data encoding. We have developed a correlation coefficient and a bit error rate to characterize the quality of a recorded pattern. Using this parameter, we can determine the optimum recording laser power. Such polarization-encoded and multilayer ODS is enabled by choosing a suitable recording and readout laser powers to digitize the TPL collected from the focus spot. This demonstrates the validity of the polarization detuning method for mitigating crosstalk through multilayer ODS. The combination of polarization-encoded ODS and multiplexing among different physical dimensions substantially increases the degrees of freedom for manipulating light-matter interactions and paves the way for the realization of green and high fidelity ODS.

## Acknowledgements
We are grateful for financial supports from the National Natural Science Foundation of China (Grant Nos. 61675093, 11674130, 91750110 and 61522504), the National Key R&D Program of China (Grant No. 2018YFB1107200), the Guangdong Provincial Innovation and Entrepreneurship Project (Grant No. 2016ZT06D081), the Natural Science Foundation of Guangdong Province, China (Grant Nos. 2016A030306016 and 2016TQ03X981), the Pearl River Nova Program of Guangzhou (Grant No. 201806010040), and the Technology Innovation and Development Plan of Yantai (Grant No. 2020XDRH095).

## Author contributions
L. W. Zhu and Y. Y. Cao contributed equally to this work. X. P. Li, and J. R. Qiu conceived the idea and designed the research; Q. Q. Chen and Z. L. Hu fabricated all the samples; L. W. Zhu, X. Ouyang and Y. Xu performed the experiments; L. W. Zhu analysed the data; L. W. Zhu and Y. Y. Cao co-wrote the manuscript. All authors discussed and commented on the manuscript.

## Competing interests
The authors declare no competing financial interests.

## Supplementary information
Supplementary information for this paper is available

# Figures:

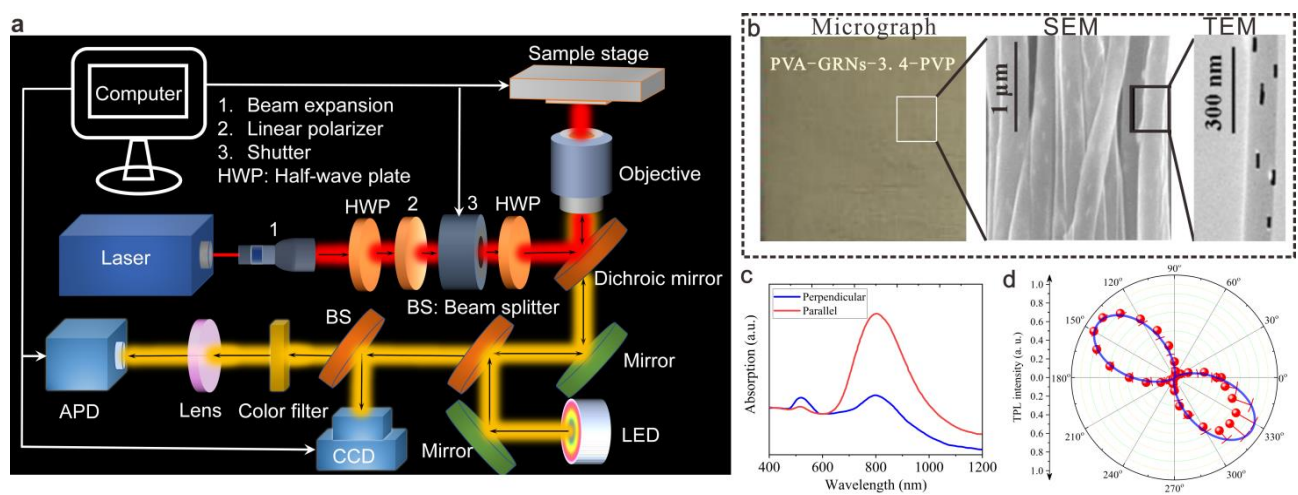

Fig. 1 (a) Schematic drawing of the optical setup for the multilayered optical data storage. (b) the optical microscope image of the data storage medium. The insets show SEM image of the aligned nanofibers and the TEM image of a single nanofiber, respectively. (c) Absorption spectrum of the nanocomposite film at different polarization of excitation light. (d) TPL polarization sensitivity of the sample excited by the corresponding wavelength. Red circles are experimental data of TPL intensities. The blue curve is the fittings with biquadratic cosine functions.

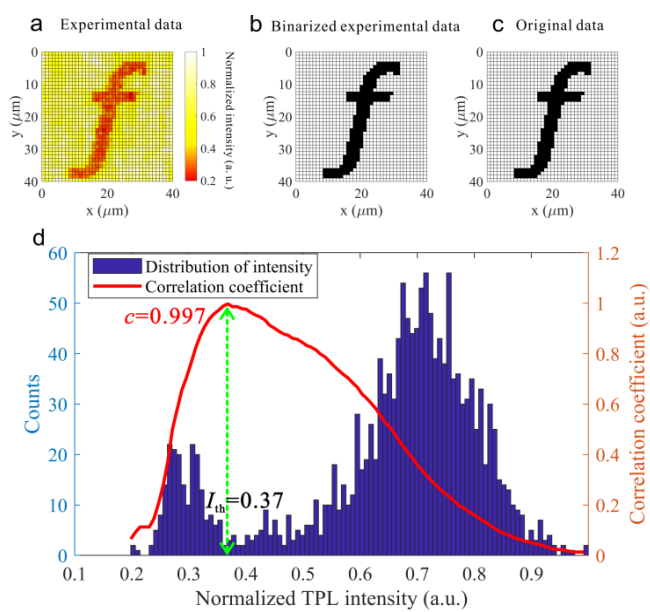

Fig. 2 (a) The pattern obtained by detecting the TPL intensities of all the information units using polarization 800nm femtosecond laser beam with the power of 50 μw. The recording powers is 1000 μw. The size of the recording region was 40 × 40 μm with 50 ×50 pixels. (b) The binarized pattern obtained by choosing an appropriate threshold intensity. (c) The original pattern used for data recording. (d) The distribution of the TPL intensities of all the information units in the extracted pattern. The calculated correlation coefficient (C = 0.997) for the extracted pattern is also provided.

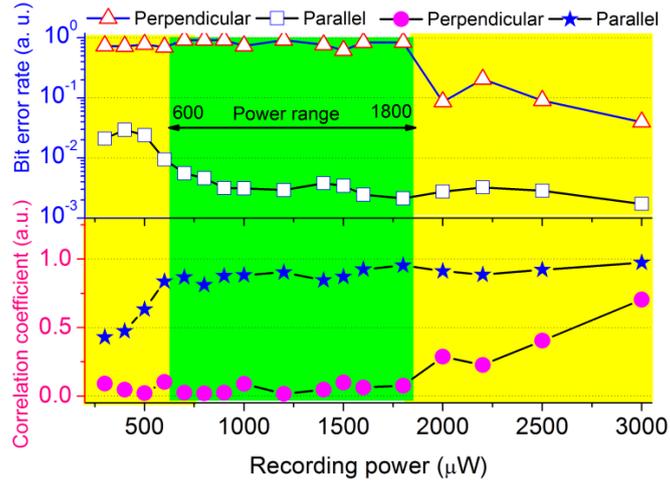

Fig. 3 Curves of the correlation coefficient and bit error rate as a function of recording power. Readout results of the TPL signals using 800 nm femtosecond laser beam with the power of 80 μw. The size of the recording region was 40 × 40 μm with 50 × 50 pixels.

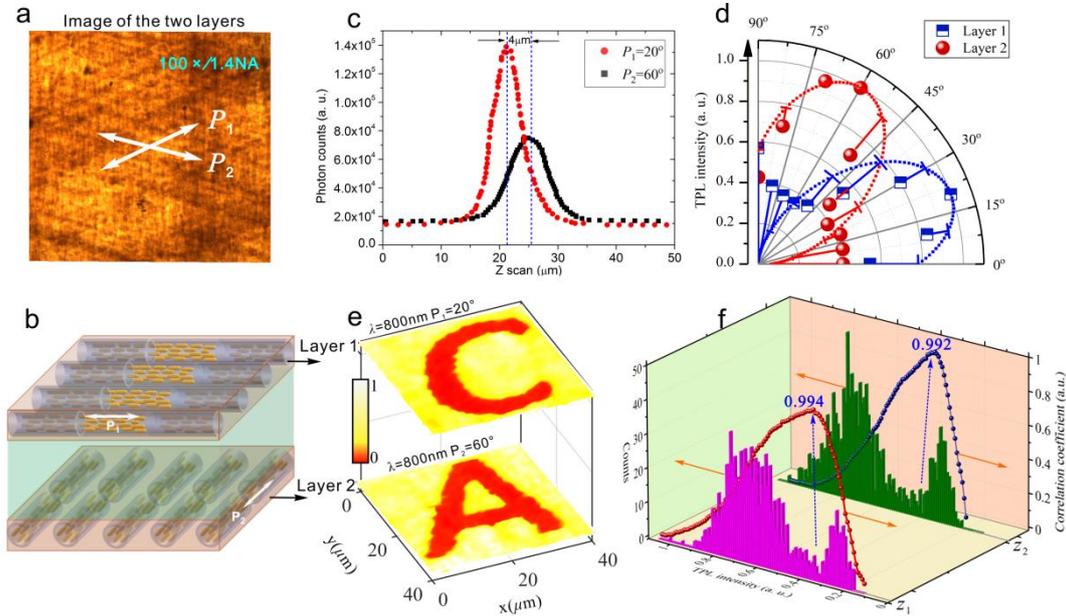

Fig. 4 (a) The optical microscope image of the two-layers data storage medium. (b) Schematic of the polarized multilayer ODS. (c) The TPL intensity signal changed as a function of Z scan depths at two polarization angles. (d) The TPL polarization sensitivity at the corresponding layers. (e) The retrieved results for the two types of combination with different layers and polarizations are shown in (b). (f) The statistical charts of the TPL intensities detected from all information units in (e). The calculated correlation coefficients for the extracted pattern are also provided.